\documentclass[article,nofootinbib,superscriptaddress,aps,prd]{revtex4-1}
\usepackage{graphicx}
\usepackage[margin=1in]{geometry}
\usepackage{subcaption}
\usepackage{amsmath}
\usepackage{amssymb}
\usepackage{color}
\usepackage[utf8]{inputenc}

\newcommand{\ben}{\begin{equation}}
\newcommand{\een}{\end{equation}}
\newcommand{\ba}{\begin{array}}
\newcommand{\ea}{\end{array}}

\def\bea{\begin{eqnarray}}
\def\eea{\end{eqnarray}}

\newcommand{\geff}{g_\text{eff}}
\newcommand{\heff}{h_\text{eff}}

\newcommand{\HubMas}{M_\text{H}}

\newcommand{\Msun}{M_\odot}
\newcommand{\Mhub}{\HubMas}
\newcommand{\Mpeak}{M_{\text{peak}}}

\newcommand{\eos}{\omega}
\newcommand{\cs}{c_\text{s}}

\def\mpc{{\rm Mpc}}

\newcommand{\deCrit}{\delta_\text{c}}
\newcommand{\de}{\delta}

\begin{document}

\title{Primordial black holes with an accurate QCD equation of state}

\author{Christian T.~Byrnes}
\email{C.Byrnes@sussex.ac.uk}
\affiliation{Department of Physics and  Astronomy, University of Sussex, Brighton BN1 9QH, UK}	

\author{Mark Hindmarsh} 
\email{m.b.hindmarsh@sussex.ac.uk}
\affiliation{Department of Physics and  Astronomy, University of Sussex, Brighton BN1 9QH, UK}
\affiliation{Department of Physics and Helsinki Institute of Physics,
		\\PL 64,  
		FI-00014 University of Helsinki,
		Finland}

\author{Sam Young}
\email{S.M.Young@sussex.ac.uk}
\affiliation{Department of Physics and  Astronomy, University of Sussex, Brighton BN1 9QH, UK}	

\author{Michael R.~S.~Hawkins}
\email{mrsh@roe.ac.uk}
\affiliation{Institute for Astronomy, University of Edinburgh, Royal Observatory, Blackford Hill, Edinburgh EH9 3HJ, UK}

\date{\today}

\begin{abstract}
Making use of definitive new lattice computations of the Standard Model thermodynamics during the quantum chromodynamic (QCD) phase transition, we calculate the enhancement in the mass distribution of primordial black holes (PBHs) due to the softening of the equation of state. We find that the enhancement peaks at approximately $0.7M_\odot$, with the formation rate increasing by at least two orders of magnitude due to the softening of the equation of state at this time, with a range of approximately $0.3M_\odot<M<1.4M_\odot$ at full width half-maximum. PBH formation is increased by a smaller amount for PBHs with masses spanning a large range, $10^{-3}M_\odot<M_{\rm PBH}<10^{3}M_\odot$, which includes the masses of the BHs that LIGO detected. The most significant source of uncertainty in the number of PBHs formed is now due to unknowns in the formation process, rather than from the phase transition. A near scale-invariant density power spectrum tuned to generate a population with mass and merger rate consistent with that detected by LIGO should also produce a much larger energy density of PBHs with solar mass. The existence of  BHs below the Chandresekhar mass limit would be a smoking gun for a primordial origin and they could arguably constitute a significant fraction of the cold dark matter density.  They also pose a challenge to inflationary model building which seek to produce the LIGO BHs without overproducing lighter PBHs.

\end{abstract}

\maketitle

\section{Introduction}

Since the paper asking ``Did LIGO detect dark matter?" appeared by Bird et al.~\cite{Bird:2016dcv} (see also \cite{Clesse:2016vqa} which appeared shortly afterwards), there has been an enormous burst of activity studying the question of whether $\sim10$ solar mass black holes, such as those which LIGO detected merging, could be primordial in origin and even make up most or all of the dark matter (DM). This activity has primarily focused on the question of whether the observational constraints on such massive black holes rule them out as being sufficiently numerous to make up all of the DM (and hence being primordial), with conflicting conclusions. Work has also been done on discriminating between a primordial or astrophysical origin of the detected black holes (BHs), for example by using their spin distribution \cite{Chiba:2017rvs} or ellipticity of the inspiralling orbit \cite{Cholis:2016kqi}, with the conclusion that future measurements should be able to make this distinction.

If the black holes detected by LIGO really are primordial, an important
question to ask is why there happens to be a peak in the primordial black hole
(PBH) number density of this mass, with fewer of both significantly
greater or smaller masses. The O($10$) solar mass range is not so far from the horizon mass at the 
quantum chromodynamic (QCD) phase transition, O(1) $\Msun$. The softening in the equation of state during this transition makes the formation probability of PBHs at around this mass scale exponentially more likely \cite{Jedamzik:1996mr}, so one is motivated to study the PBH 
mass distribution across an extended mass range.

In this paper, we quantify the effect on the PBH mass distribution due to the QCD transition. 
While the effect has been studied before, see e.g.~\cite{Jedamzik:1998hc,Jedamzik:1999am,Widerin:1998my,Sobrinho:2016fay}, new lattice results 
\cite{Borsanyi:2016ksw,Bhattacharya:2014ara} now provide a definitive equation of state of the universe through the QCD phase transition, 
allowing for the first time an exploration of black hole formation at this epoch with negligible thermodynamic uncertainties.

We find that the equation of state parameter $\omega=p/\rho$ reduces by around $30\,\%$ during the QCD phase transition, and that the corresponding decrease in the critical collapse threshold of the comoving density contrast is around $10\,\%$. 
This leads to a boost in the PBH mass distribution 
by at least two orders of magnitude compared to a Universe in which the equation of state 
parameter remained $\omega=1/3$. The peak spans about one order of magnitude in the range of black hole masses, with the peak at around one solar mass, but the 
enhancement can be seen in the even wider mass range of $ 10^{-3} \lesssim M/\Msun \lesssim 10^{3}$. 
The enhancement in the density of BHs in the mass range detected by LIGO is around a factor of 5; 
$30\Msun$ black holes are forming at around the time muons and pions become non-relativistic.
Our results are not very sensitive to different assumptions about the formation criteria of PBHs with a time varying equation of state and we discuss the various choices, while stressing the need for numerical simulations in order to reach a precise prediction.

Our results show that if a near-scale-invariant density power spectrum 
is tuned to generate PBHs with a density at $M\sim10 M_\odot$ to match the inferred LIGO merger rate \cite{Sasaki:2016jop,Ali-Haimoud:2017rtz}
then there should also exist a much larger density of PBHs with solar mass, unless the power spectrum is very red on the relevant scales. 
Solar mass black holes are below the Tolman--Oppenheimer--Volkoff and Chandrasekhar mass limits and hence, if detected, would be strong evidence that the black holes are primordial.

A corollary is that it is possible that PBHs are responsible both for the population of black holes observed by LIGO 
and for a significant fraction of the dark matter. There is evidence for such a dark matter population from microlensing of quasar light, which indicates that 
most, if not all, quasars are microlensed by compact bodies of around a solar mass \cite{m09,h11,p73},  see also \cite{Mediavilla:2017bok} which argues this microlensing is consistent with the expected stellar population. For example, a locally scale-invariant density power spectrum with variance $\sigma^2=0.004$ gives    
PBHs making up 13\% of DM with in the mass range $0.2 < M/\Msun < 1$, 
and with 0.1\% in the range $10 < M/\Msun < 50$.  
  A population of PBHs much greater than a solar
mass appears to be ruled out by several dynamical and accretion
constraints, although there is significant uncertainty in the assumptions
which these methods typically entail.  The mass range from
$10^{-5} M_{\odot}$ to $5 M_{\odot}$ is primarily constrained by microlensing
observations \citep{Carr:2017jsz} which depend on a knowledge of the structure of
the Galactic halo \citep{Hawkins:2015uja}.  We discuss these constraints in more
detail in Section V.  
See \cite{Sasaki:2018dmp} for a recent review of all observational constraints.

The plan of this paper is as follows: in Sec.~\ref{sec:eos} the recent results for the equation of state near the QCD phase transition 
are reviewed. 
We use those results in Sec.~\ref{sec:deltac} to discuss the reduction of the critical collapse threshold during the phase transition. In Sec.~\ref{sec:mass-function} we derive the mass function of PBHs expected during the phase transition, and discuss how this varies if the underlying power spectrum is not scale-invariant over the relevant range of scales. The observational evidence for and against solar mass PBHs from different sources is discussed in Appendix~\ref{sec:observations} and we conclude in Sec.~\ref{sec:conclusions}.

\section{Equation of state during the QCD phase transition}\label{sec:eos}

As the universe cooled through a temperature of order 100 MeV, the effective number of relativistic degrees of freedom changed 
rapidly as the strong interactions confined quarks into hadrons.  
Lattice QCD studies \cite{Bazavov:2014pvz,Borsanyi:2013bia} agree that the transition is a cross-over, which means that thermodynamic quantities evolve smoothly, and there is no period of phase coexistence (as in a first-order transition) or divergent specific heat (as in a second-order transition).
Instead, the equation of state parameter $\eos(T)$ 
and the sound speed squared $\cs^2(T)$ 
dip well below $1/3$ at around this temperature, but suffer no discontinuity.

We denote the two relevant measures of the effective number of relativistic degrees of freedom as 
\ben
\geff(T) = \frac{30 \rho}{\pi^2 T^4}, \quad \heff(T) = \frac{45 s}{2\pi^2T^3}, 
\een
where $\rho$ is the energy density and $s$ the entropy density. From the relationship between the energy density, entropy density and the pressure $p = sT - \rho$, we see that the equation of state parameter is given by
\ben
\eos(T) = \frac{4\heff(T)}{3\geff(T)} - 1
\een
and the sound speed squared by
\ben
\cs^2(T) = \frac{4(4\heff + T\heff'(T))}{3(4\geff + T\geff'(T))} - 1,
\een
where the prime indicates differentiation with respect to temperature.

The uncertainties in the functions $\geff(T)$ and $\heff(T)$ near the QCD transition have reduced dramatically recently as improved computing power and techniques have enabled the use of quarks with physical masses to study the equation of state of QCD at finite temperature.  We will use the results of Ref.~\cite{Borsanyi:2016ksw}, which combine lattice results near the transition with hard thermal loop effective theory at high temperatures and the hadronic resonance gas at low temperature, to produce a definitive equation of state for the Standard Model for temperatures in the range $1 \lesssim T \lesssim 10^5$ MeV. 
This is the first time an accurate Standard Model equation of state has been used in studies of PBH formation.

\begin{figure}[t]
\includegraphics[width=0.7 \textwidth]{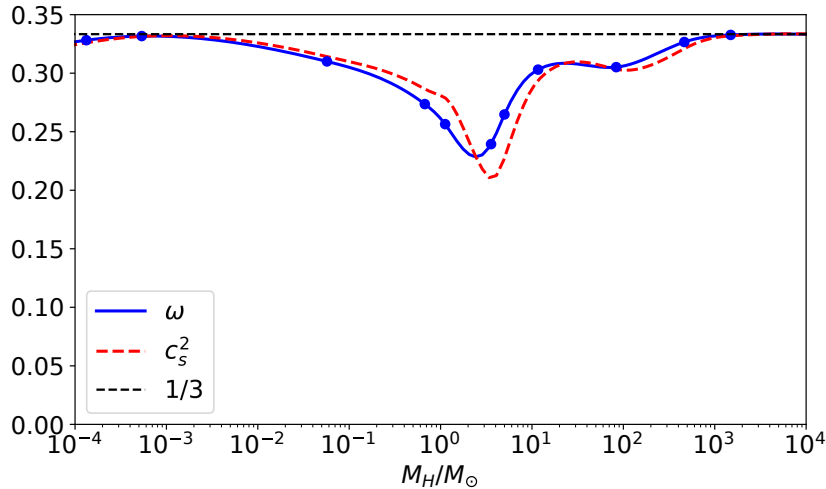}
\caption{The equation of state parameter $\eos$ and the sound speed squared $\cs^2$ for the Standard Model \cite{Borsanyi:2016ksw}, 
plotted against horizon mass, in units of solar mass. The filled circles show the nodes of the spline used for the fit for $\eos$. 
The horizon mass range plotted corresponds to a temperature range of approximately $10^4$~MeV to $1$~MeV.
\label{EoSmass} }
\end{figure}

Ref.~\cite{Borsanyi:2016ksw} tabulates values of $\geff(T)$ and $\geff(T)/\heff(T)$ in a form suitable for spline interpolation, 
giving about 1\% accuracy in $\geff$ and $0.3\%$ for the ratio.
In Figure \ref{EoSmass} we show the resulting equation of state parameter $\eos(T)$ 
and speed of sound squared $c_s^2(T)$, 
plotted against the horizon mass\footnote{More correctly but less commonly called the Hubble volume mass.} $\Mhub$ in units of the solar mass $\Msun$. 
It can be seen that, although these quantities are smooth, there is a distinct minimum where the Hubble volume mass is O(1) $\Msun$, 
corresponding to the temperatures around 200 MeV.  
The dip is not as extreme as in the models of the transition used in Ref.~\cite{Sobrinho:2016fay},
or in previous models of the Standard Model equation of state based on earlier lattice data  \cite{Hindmarsh:2005ix,Laine:2006cp}.

In fact, there are significant departures from pure radiation 
thermodynamics across a wide range of scales. 
At temperatures above the QCD phase transition, they are due to quark mass thresholds and non-perturbative effects in the non-Abelian gauge fields.
There is also a broad dip at temperatures just below the transition, corresponding to horizon mass range $10$ to $100$  $\Msun$, 
due to pions and muons becoming non-relativistic, resulting in the disappearance of $13/2$ of the remaining $69/4$ relativistic degrees of freedom. 
This is also an interesting range, as it encompasses the masses of the merging binaries detected by LIGO.

\section{Critical density perturbation for collapse during the phase transition}\label{sec:deltac}

PBHs may have formed in the early Universe shortly after inflation ended from the direct collapse of large density perturbations (for reviews see \cite{Carr:2009jm,Green:2014faa,Sasaki:2018dmp}). When a density perturbation reenters the horizon, it will collapse to form a PBH if the density contrast is above some critical threshold, $\delta_c$. 

Musco and Miller \cite{Musco:2012au} simulated the formation of PBHs with differing values for the equation of state, calculating the critical density perturbation for collapse $\deCrit$ as function of the equation of state parameter $\eos$, and these results are used in this paper. However, it should be noted that these results were calculated assuming a constant equation of state. The true critical value therefore remains uncertain. 
However, the general tendency is clear: a decrease in $\eos$ results in a decrease in $\deCrit$, as shown in figure 8 of Ref.~\cite{Musco:2012au}.
Therefore the reduced equation of state parameter during the phase transition results in the enhanced formation of PBHs. 

There are two key questions to consider:
at which mass does the peak in the PBH distribution occur, 
and how high is the peak relative to the value it would take if there were no phase transition.

In order to investigate the peak mass, the critical density perturbation will be calculated from the equation of state parameter at two key times during the formation of a PBH: 1) the time when the perturbation first enters the horizon and the process of collapse begins, and 2) the time when the PBH forming region stops expanding and starts to collapse. This will give a minimum and maximum for the peak formation time and a corresponding peak mass. The true value for the peak formation time will lie somewhere between these values.

The height of the peak will depend on how much the critical density perturbation changes during the transition, which will depend on the equation of state throughout the entire formation process. To fully answer this will require further investigation, although for the purpose of this paper, an averaged value of the equation of state parameter during the PBH formation time is used. Both a linear and logarithmic time average 
are considered; the resulting mass spectrum is found to be insensitive to the averaging method.

First we recall the time at which a forming PBH forming stops expanding, the turn-around time. 
Consider a spherical density perturbation of amplitude $\delta_*$ at time $t_*$ on a spatially flat radiation-dominated Friedmann universe with background density $\rho_{*c}$ and suppose that the overdense region can be considered as a separate uniform-density closed universe. For the overdense region, the Hubble parameter $H$ is given by the Friedmann equation,
\begin{equation}
H^2=\frac{8\pi G}{3}\rho-\frac{Kc^2}{a^2},
\end{equation}
where $a$ is the scale factor, $K$ is the curvature, $G$ is the gravitational constant, $\rho$ is the density of the overdense region, and $c$ is the speed of light. The comoving density contrast at an initial time $t_*$ is given by
\begin{equation}
\delta_*=\frac{\rho_*}{\rho_{*c}}-1=\frac{K}{a_*^2H_*^2}.
\label{eqn:dencon}
\end{equation}

The overdense region will eventually cease expanding and start to re-collapse. The time at which this happens is called the turn-around time, and within the simple model being considered, if this time is reached the subsequent formation of a PBH is inevitable. At the turn-around time $H_{\rm ta}=0$, so the density satisfies
\begin{equation}
\frac{8\pi G}{3}\rho_{\rm ta}=\frac{K}{a_{\rm ta}^2}
\end{equation}
where the subscript ``ta'' denotes the turn-around time. The ratio between the energy density at an initial time $t_*$ and turn-around time $t_{\rm ta}$ is given by
\begin{equation}
\frac{\rho_{\rm ta}}{\rho_*}=\frac{K}{a_*^2H_*^2+K}\frac{a_*^2}{a_{\rm ta}^2}.
\end{equation}
Using equation (\ref{eqn:dencon}), this can be rewritten as
\begin{equation}
\frac{\rho_{\rm ta}}{\rho_*}=\frac{\delta_*}{1+\delta_*}\frac{a_*^2}{a_{\rm ta}^2}.
\end{equation}
Assuming a constant equation of state (during the phase transition, the equation of state does not change greatly  from 1/3, see Fig.~\ref{EoSmass}), the energy density evolves as
\begin{equation}
\rho_{\rm ta}=\left( \frac{a_{\rm ta}}{a_*} \right)^{-3(1+\omega)}\rho_*.
\end{equation}
Combining these equations gives the ratio between the initial scale factor and the scale factor at the turn-around time,
\begin{equation}
\frac{a_{\rm ta}}{a_*}=\left( \frac{1+\delta_*}{\delta_*} \right)^{\frac{1}{1+3\omega}}.
\end{equation}
The initial time will be taken as the time the perturbation enters the horizon, and at this time $\delta_*$ will be taken as the minimum value required for a PBH to form during radiation domination, $\delta_*=\delta_c=0.453$ \cite{Musco:2012au},
The number of efolds to reach turn around is therefore
\bea
N_{\rm ta}\simeq\frac{1}{2} \ln(1.453/0.453)\simeq 0.6.
\label{Nta}
\eea
During this time, the horizon mass grows by a factor of $\exp(2 N_{\rm ta})\approx 3$. 

\begin{figure}
\includegraphics[width=0.7 \textwidth]{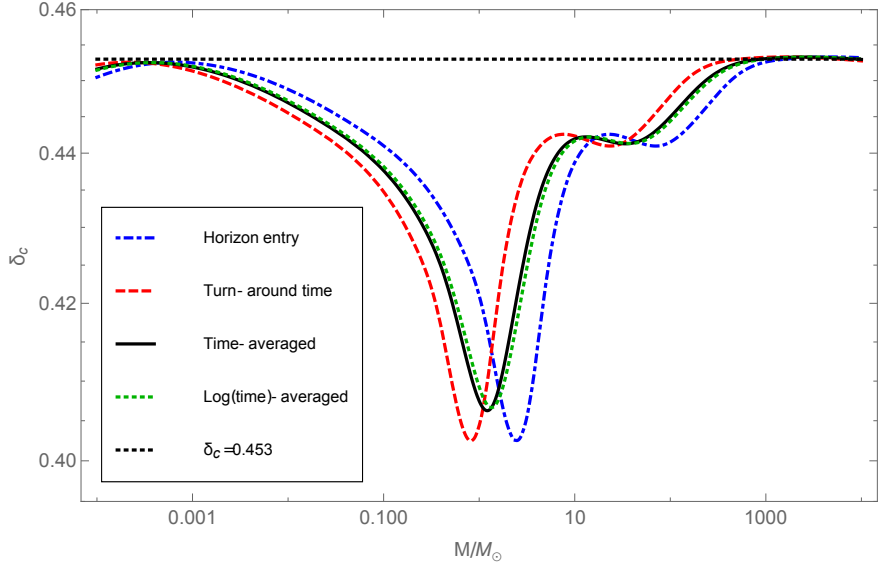}
\caption{The critical density perturbation for collapse $\delta_c$ as a function of horizon mass during the QCD phase transition. 
The critical value is calculated from the equation of state at horizon entry (dot-dashed blue line), turn-around (dashed red line), and two time-averaged values, with linear (solid black line) and logarithmic (dotted green line) averaging. 
The critical density perturbation for a pure radiation fluid $\delta_c=0.453$ is also plotted with a dashed black line for reference.}
\label{CriticalVsMass} 
\end{figure}

Figure \ref{CriticalVsMass} shows the evolution of the critical density perturbation $\delta_c$ as a function of horizon mass through the phase transition, with four different methods of treating the equation of state parameter during the collapse; taking its value at horizon-entry, turn-around time, and two different time averages. 
It can be seen that averaging the equation of state increases the minimum value for $\delta_c$, and that there is a negligible difference between linear and logarithmic time averaging.

\section{The mass function of primordial black holes}\label{sec:mass-function}

Typically, the PBH abundance is stated in terms of the mass fraction of the Universe that collapses to form PBHs at the time of formation $\beta$, and the simplest version of the calculation unrealistically assumes that all PBHs form with exactly the horizon mass. In this case, a Press-Schechter approach is used to calculate $\beta$
\begin{equation}
\beta=2\int\limits_{\delta_c}^{\infty}d\delta P(\delta),
\end{equation}
where $\delta_c$ is the critical value for PBH formation at the time of horizon entry, and $P(\delta)$ is the probability density function of the density contrast (and we have multiplied the above by the Press-Schechter factor of two \cite{Press:1973iz}). In this paper, a Gaussian distribution for $P$ will be assumed, although it should be noted that even small amounts of non-Gaussianity can have a significant effect \cite{Bullock:1996at,Ivanov:1997ia,PinaAvelino:2005rm,Byrnes:2012yx,Shandera:2012ke,Young:2013oia,Young:2015cyn}. 

We write the Gaussian probability distribution as 
\begin{equation}
P(\delta)=\frac{1}{\sqrt{2\pi \sigma^2}}\exp\left( -\frac{\delta^2}{2\sigma^2} \right),
\end{equation}
where the variance of the density perturbations, $\sigma^2$, is calculated by integrating the density power spectrum $\mathcal{P}_\delta(k)$ \cite{Young:2014ana}, 
\begin{equation}
\sigma^2=\int\limits_{0}^{\infty}W^2(kR)\mathcal{P}_\delta(k)\frac{\mathrm{d}k}{k},
\end{equation}
where $W(kR)$ is the Fourier transform of the window smoothing function, and $R$ is the horizon scale at a given time. For a Gaussian distribution, $\beta$ can be approximated as
\begin{equation}
\beta=\mathrm{erfc}\left( \frac{\delta_c}{\sqrt{2 \sigma^2}} \right).
\label{eq:beta}
\end{equation}
In this paper, the power spectrum amplitude at PBH formation scales is a free parameter that may be tuned to produce the correct number of PBHs to make up dark matter. We first assume for simplicity that the density power spectrum at horizon-entry is scale-invariant over a relevant but limited range of scales, leading to a constant $\beta$ on those scales if the critical value for collapse is constant.

However, whilst a scale-invariant power spectrum leads to a scale invariant $\beta$, it does not lead to a scale invariant PBH mass spectrum. This is due to the fact that, once formed, the number density of PBHs evolves like matter during radiation domination. The radiation density of the Universe evolves as 
$\rho_{\rm rad}\propto a^{-4}$, whilst the matter density evolves as $\rho_{\rm mat}\propto a^{-3}$. This means that the number density of PBHs grows proportional to the scale factor during radiation domination. Assuming that black holes form with mass $M = \Mhub$, and that 
the Universe behaves purely as radiation domination until the time of matter-radiation equality means that the abundance of PBHs at that time $\beta_{\rm eq}$ can be stated in terms of $\beta$ as \begin{equation}
\beta_{\rm eq}(M)=\frac{a_{\rm eq}}{a(M)}\beta(M)=\left( \frac{M_{\rm eq}}{M}\right)^{1/2}\beta(M),
\end{equation}
where the subscript ``eq'' denotes the scale factor at the time of matter-radiation equality, and $a(M)$ is the scale factor  
when the horizon mass is $M$.  
The final abundance of PBHs therefore decreases with increasing mass. The horizon mass at the time of matter-radiation equality is $M_{H,{\rm eq}}\approx 2.8\times 10^{17}M_\odot$ (using the same cosmological parameters as Ref.~\cite{Nakama:2016gzw}) 
The fraction $\beta_{\rm eq}$ represents the abundance at the time of matter-radiation equality of PBHs which formed at a given earlier time - in order to calculate the total abundance of PBHs, $\Omega_{\rm PBH}$, it is necessary to integrate $\beta_{\rm eq}$ over all times at which PBHs form, 
\begin{equation}
\Omega_{\rm PBH}=\int\limits_{M_{\rm min}}^{M_{\rm max}}\mathrm{d}\ln(M)\beta_{\rm eq}(M)=\int\limits_{M_{\rm min}}^{M_{\rm max}}\mathrm{d}\ln(M)\left( \frac{M_{\rm eq}}{M_{\rm PBH}}\right)^{1/2}\beta(M).
\end{equation}
The mass distribution of PBHs will be stated in terms of $f(M)$, the fraction of CDM made up of PBHs of a given mass $M$,
\begin{equation}
f(M)=\frac{1}{\Omega_{\rm CDM}}\frac{\mathrm{d}\Omega_{\rm PBH}}{\mathrm{d}\ln{M}}.
\label{eqn:f}
\end{equation}
In the simplified case currently being considered, if PBHs formed with exactly the horizon mass, the expression for $f$ measured after matter-radiation equality would be
\begin{equation}
f=\left( \frac{M}{M_{\rm eq}}\right)^{-1/2}\frac{\beta(M)}{\Omega_{\rm CDM}}.
\end{equation}
In the next subsection we perform a more realistic calculation by including the effects of critical collapse on the PBH mass spectrum.

For reference we also give approximate relations between the horizon mass with time, wavenumber and scale factor  \cite{Carr:2009jm,Nakama:2016gzw}
\bea
M_H&\simeq& 2\times10^5 \left(\frac{t}{1{\rm s}}\right) M_\odot \\ 
&\simeq& 1.5\times10^5 \left(\frac{g}{10.75}\right)^{-1/2}\left(\frac{T}{1{\rm MeV}}\right)^{-2} M_\odot \\
&\simeq& 17 \left(\frac{g}{10.75}\right)^{-1/6}\left(\frac{k}{1{0^6\rm Mpc}^{-1}}\right)^{-2} M_\odot \label{M-k}
\eea
where $g$ is the number of degrees of freedom of relativistic particles.

\subsection{Extended mass function of primordial black hole formation}
The calculation will now be extended to account for the fact that PBHs do not form with exactly the horizon mass. 
The mass of the resulting PBH depends on the horizon mass and the amplitude of the density perturbation $\delta$ defined in the comoving gauge from which it formed, and is given by \cite{Niemeyer:1997mt,Musco:2004ak,Musco:2008hv,Musco:2012au}
\begin{equation}
M=k \Mhub (\delta-\delta_c)^\gamma,
\end{equation}
where $\Mhub$ is the horizon mass at the time of horizon re-entry, and during radiation domination the constants have been numerically found to be given by $k=3.3,\;\gamma=0.36,\;\delta_c=0.453$ (the values depend upon the radial profile of the perturbations being considered; we use the values given here in order to be concrete). We assume that only $\delta_c$ varies with the equation of state. This equation can be inverted to give $M$ as a function of $\delta$,
\begin{equation}
\delta=\left( \frac{M}{k\Mhub} \right)^{1/\gamma}+\delta_c.
\end{equation}
The expression for $\beta$ is therefore amended to account for the fraction of each Hubble volume which collapses to form a PBH:
\begin{equation}
\beta=2\int\limits_{\delta_c}^{\infty}  \frac{M}{\Mhub}P(\delta) d\delta=2\int\limits_{\delta_c}^{\infty} k (\delta-\delta_c)^\gamma P(\delta) d\delta.
\end{equation}
The final expression expression for $f$ can be found by combining the previous two equations with equation (\ref{eqn:f}), giving $f$ as a function of the PBH mass $M$ and $\sigma^2$ \cite{Niemeyer:1997mt}, 
\begin{equation}
f(M)= \frac{1}{\Omega_{\rm CDM}} \int\limits_{-\infty}^{\infty} \frac{2}{\sqrt{2\pi\sigma^2(\Mhub)}} 
\exp\left[{-\frac{(\mu^{1/\gamma}+\delta_c(\Mhub))^2}{2\sigma^2(\Mhub)}} \right]
\frac{M}{\gamma \Mhub}\mu^{1/\gamma}\sqrt{\frac{M_{\rm eq}}{\Mhub}} d\ln \Mhub,
\end{equation}
where $\mu\equiv\frac{M}{k \Mhub}$ and we have used $\frac{d\delta}{d\ln M}=\frac{1}{\gamma}\mu^{1/\gamma}$.
The difference which critical collapse behaviour makes is to reduce the mean black hole mass below $\Mhub$ and to broaden the distribution 
\cite{Yokoyama:1998xd} (see also \cite{Kuhnel:2015vtw}).
The total contribution of PBHs today is given by
\begin{equation}
\Omega_{\rm PBH}=\Omega_{\rm CDM} \int f(M) d\ln M,
 \end{equation}
 where the limits of the integral should include all masses of PBHs which form.
 
\begin{figure}
\begin{center}

\begin{tabular}{cc}
    \includegraphics[width=0.5\linewidth]{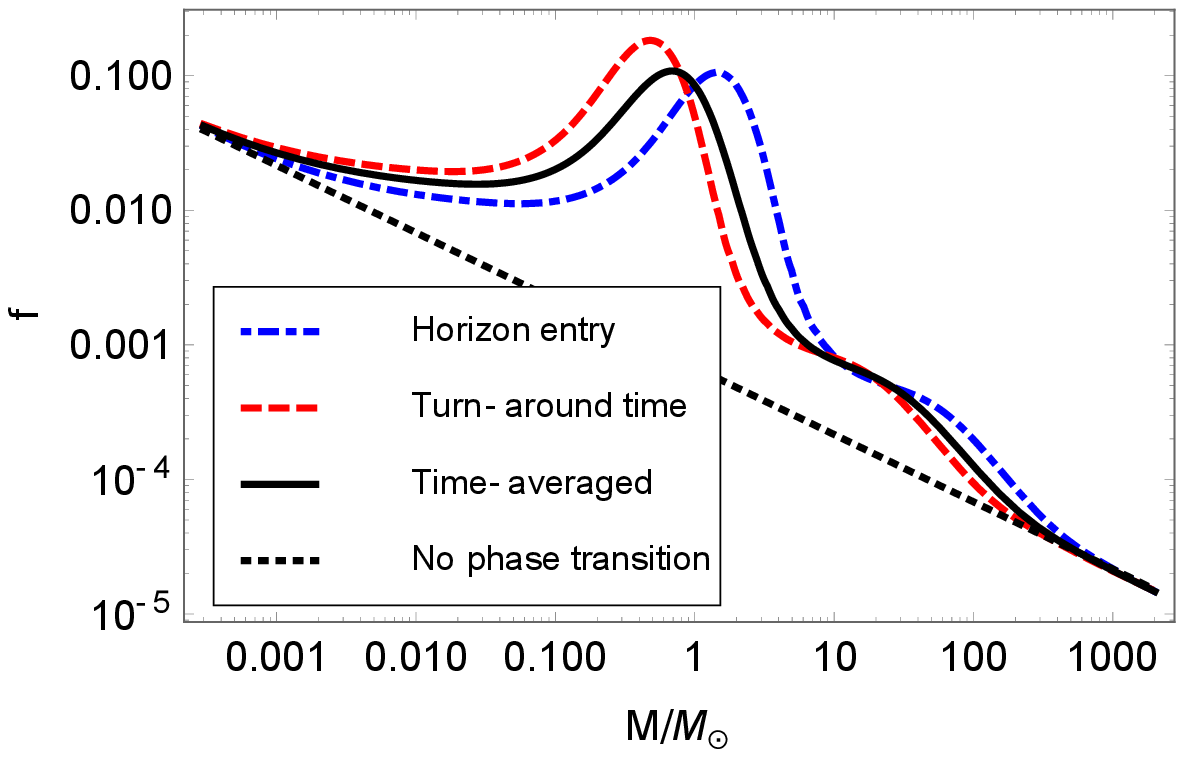}&
    \includegraphics[width=0.5\linewidth]{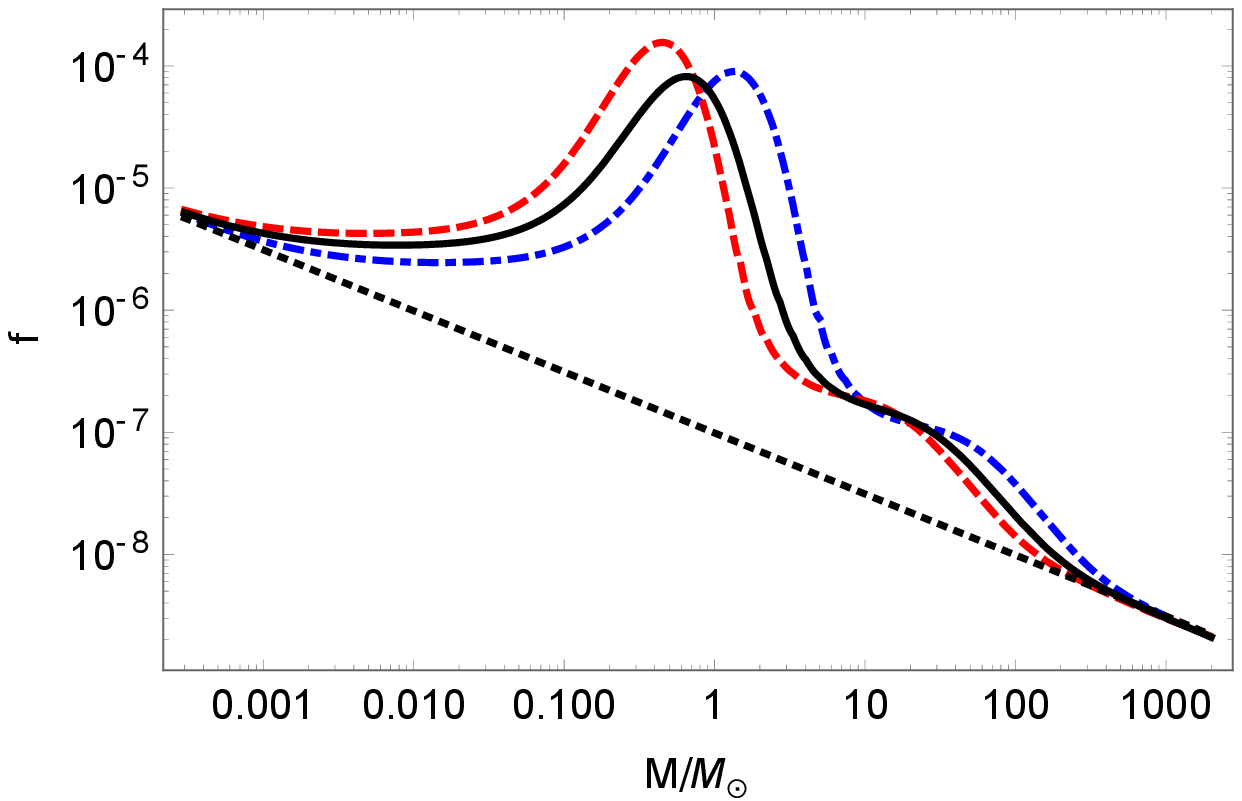}\\
\end{tabular}

\end{center}
\vspace*{-0.5cm}
\caption{The mass distribution $f$ of PBHs forming during the phase transition is shown 
for a scale-invariant density power spectrum, with different ways of treating the equation of state parameter during the black hole formation process. The different lines correspond to using the equation of state at horizon entry, turn-around and a time-averaged value. The straight dashed black line represents the mass function $f$ of PBHs if there is no phase transition (taking the critical density perturbation $\delta_c=0.453$). The variance of the density contrast 
at horizon crossing 
is taken to be $\sigma^2=0.004$ for the left plot and $\sigma^2=0.003$ for the right. Using the time-averaged value of $\delta_c$, for  $\sigma^2=0.004$, the peak occurs at $0.69M_\odot$ and the range at half-maximum is $0.30M_\odot<M<1.4M_\odot$, whilst for  $\sigma^2=0.003$, the peak (which becomes sharper for smaller values of $\sigma^2$) occurs at $0.65M_\odot$ and the range at half-maximum is $0.32M_\odot<M<1.2M_\odot$.
\label{f-004-full}} 
\end{figure}

The left hand plot in Fig.~\ref{f-004-full} plots $f(M)$ for $\sigma^2=0.004$ from $2\times10^{-4}M_\odot$ to $3\times10^3 M_\odot$ (which corresponds to $N=\ln(M_{\rm max}/M_{\rm min})/2 \simeq 8$ efolds). 

Since PBHs form from large density perturbations, in the tail of distribution, their abundance is exponential sensitive to the variance of the density
perturbations (\cite{Carr:1975qj}, c.f.~Eq.~\eqref{eq:beta}). In Fig. 3 we see that decreasing the power spectrum amplitude by 25$\%$ reduces the PBH energy density
by several orders of magnitude.
Hence only a relatively narrow range of $\sigma^2$ is of observational interest, as a result of the exponential dependence of $\beta$ on the amplitude of the primordial power spectrum. We note that the position of the peaks does not change significantly when changing the amplitude of the power spectrum, but that the sharpness of the peaks does increase as $\sigma^2$ is decreased, growing from two orders of magnitude above the ``background'' value (i.e.~the value of $f$ one would calculate assuming a pure radiation dominated background) to an enhancement by three orders of magnitude over the background value.

It can be seen that calculating the critical value for collapse from the equation of state at horizon entry, turn-around, or an averaged value leads to about a factor of two change in the peak mass at which PBHs form, and a smaller change in the height of the peak. For the rest of this paper we will use the time averaged value of $\delta_c$, defined by  
\ben
\label{e:TimAveDelC}
\bar\delta_c(\Mhub) = \left\{ \ba{cl}
\displaystyle \frac{1}{t_\text{ta} - t_\text{H}} \int_{t_\text{H}}^{t_\text{ta}} dt \de_c, & \textrm{time average} \\
\displaystyle \frac{1}{\ln\left({ t_\text{ta}}/{t_\text{H}}\right) } \int_{t_\text{H}}^{t_\text{ta}} \frac{dt}{t} \de_c, & \textrm{logarithmic time average} 
\ea
\right.
\een
This value is likely to be closest to the true answer, since the PBH formation will be sensitive to the equation of state during the complete period between horizon entry and when the overdensity stops expanding. Fig.~\ref{CriticalVsMass} shows that  $\delta_c$ is quite insensitive to the averaging procedure used. Following Eq.~\eqref{Nta} we use $t_{\rm ta}= 3 t_{\rm H}$ in order to be concrete.

\begin{table}

\begin{tabular}{l | c | c | c }
\hline
 & $M_{\text{fwhm}}^\text{low}/\Msun$ & $\Mpeak/\Msun$ & $M_{\text{fwhm}}^\text{high}/\Msun$ \\
 \hline
 horizon entry & 0.69 & 1.4 & 2.5 \\
 turn-around & 0.23 & 0.48 & 0.83 \\
 averaged & 0.30 & 0.69 & 1.4 \\
\hline
\end{tabular}
\caption{\label{t:MassDist}
Parameters of the PBH mass distribution (peak mass and upper and lower values for the full-width half-maximum) 
for a scale-invariant primordial density perturbation spectrum with 
$\sigma^2 = 0.004$, with different assumptions about how the critical collapse threshold depends on the 
equation of state. Linear and logarithmic time averaging are almost indistinguishable at this level of precision.
}

\end{table}

\subsection{Implications for the LIGO detection of intermediate mass BHs}

Since the LIGO detection of several in-spiralling intermediate mass black holes \cite{Abbott:2017gyy} 
there has been great interest in whether the BHs which LIGO detected were primordial or astrophysical, 
e.g.~\cite{Bird:2016dcv,Clesse:2016vqa,Raccanelli:2016cud}. 
There is a significant debate about whether 
PBHs  in the mass range of $10-50 M_\odot$ detected by LIGO could make up all of the DM, see e.g.~\cite{Carr:2016drx,Bernal:2017vvn}, with no definitive conclusion reached due to various uncertainties in both the constraints and the expected properties of PBHS, see e.g.~\cite{Green:2016xgy,Nakama:2016gzw,Green:2017qoa,Carr:2017jsz,Garcia-Bellido:2017xvr,Clesse:2017bsw,Zumalacarregui:2017qqd,Garcia-Bellido:2017imq,Bernal:2017vvn,Bellomo:2017zsr,Lehmann:2018ejc} and Appenix \ref{sec:observations}. 
However, there are recent works estimating that the merger rate of PBHs would match the LIGO detection rate 
if around $0.1\%$ of DM is made up of PBHs in the LIGO mass range 
\cite{Sasaki:2016jop,Ali-Haimoud:2017rtz,Raidal:2017mfl,Kocsis:2017yty} (assuming that the PBHs are not initially clustered). 

\begin{figure*}[tb]
\centering
\begin{tabular}{ccc}
    \includegraphics[width=0.33\linewidth]{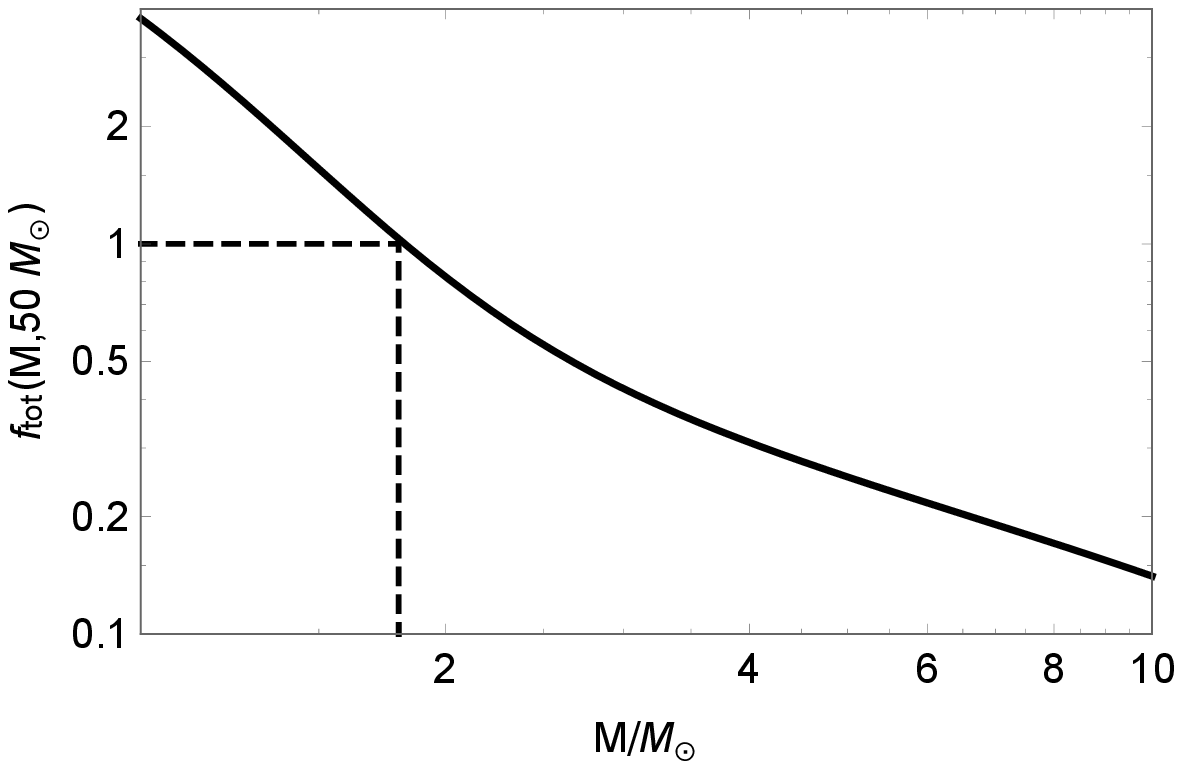}&
    \includegraphics[width=0.33\linewidth]{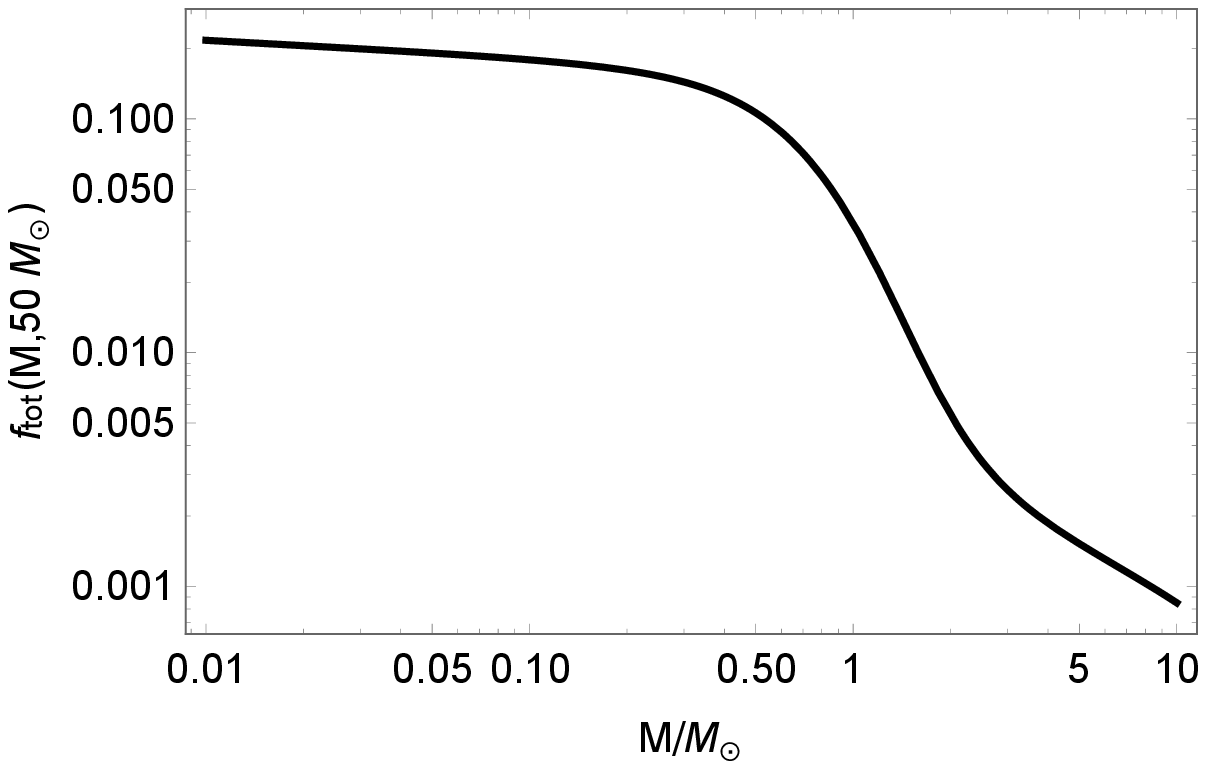}&
    \includegraphics[width=0.33\linewidth]{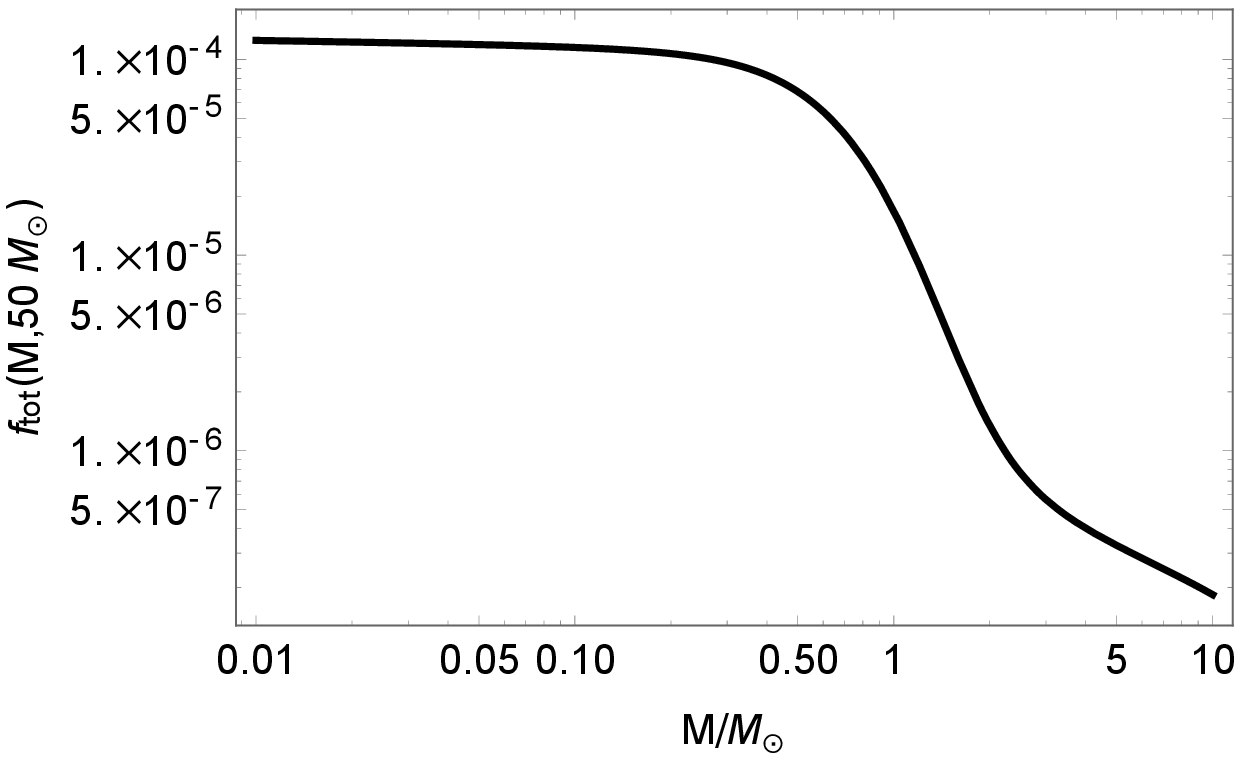}\\
\end{tabular}
\caption{\label{fig:f-tot} The total mass fraction in PBHs between the mass plotted on the $x$-axis and $50 M_\odot$, as defined by Eq.~\eqref{f-tot-defn}, using the time-averaged equation of state. The amplitude of the power spectrum changes from $\sigma^2=0.005$ on the left to $\sigma^2=0.004$ in the middle and $\sigma^2=0.003$ on the right plot and the fraction of DM in PBHs with masses between $10-50M_\odot$ 
(i.e.~the value of $f_{\rm tot}(10 M_\odot,50 M_\odot)$, Eq.~\ref{f-tot-defn}) is approximately $0.14, 0.001$ and 
$2\times10^{-7}$ 
from left to right respectively. The PBH density exceeds the measured DM density if the scale invariant spectrum is assumed to extend below $1.8 M_\odot$, and therefore the scale invariance cannot extend below this mass. }
\end{figure*}

Based on  the above  discussion, 
we define the integrated fraction of DM between masses $M_1$ and $M_2$ as 
\begin{equation}
f_{\rm tot}(M_1,M_2) = \int^{M_2}_{M_1} f(M) \frac{{\rm d} M}{M}.
\label{f-tot-defn}
\end{equation}
We plot $f_{\rm tot}(M,50 M_\odot)$ for three cases: $\sigma^2 = 0.005$, $0.004$, and $0.003$ in Fig.~\ref{fig:f-tot}, using 
the time-averaged $\delta_c$ (Eq.~\ref{e:TimAveDelC}, corresponding to the dashed black line in Fig.~\ref{CriticalVsMass}). 
For comparison, the integrated fraction  $f_{\rm tot}(10 M_\odot,50 M_\odot)=0.15$, $0.001$ and $2\times10^{-7}$, respectively. 
Notice how the value of $f_{\rm tot}$ doubles on a much shorter scale (in a logarithmic sense, which corresponds to efolding number during inflation) towards lighter PBH masses, especially for smaller amplitudes of the power spectrum, due to the spike in the energy density of PBHs with mass close to one solar mass. For example, for the middle plot, $f_{\rm tot}(2M_\odot,10M_\odot)\simeq 10 f_{\rm tot}(10M_\odot,50M_\odot)$, despite the mass ranges spanning an equal duration of inflation when measured in efolding number.


\subsection{Scale-dependent power spectra}

For a strictly scale invariant power spectrum, as used in 
all plots of the PBH mass function shown thus far, 
the mass distribution diverges at small masses. 
To remove the divergence 
a sufficiently red tilt can be introduced to the primordial curvature perturbation spectrum. 
Because of the exponential sensitivity of the mass fraction to the variance $\sigma^2(M)$, a small tilt changes the slope of the mass distribution a lot.

\begin{figure*}[tb]
\centering
\begin{tabular}{cc}
\includegraphics[width=0.5\linewidth]{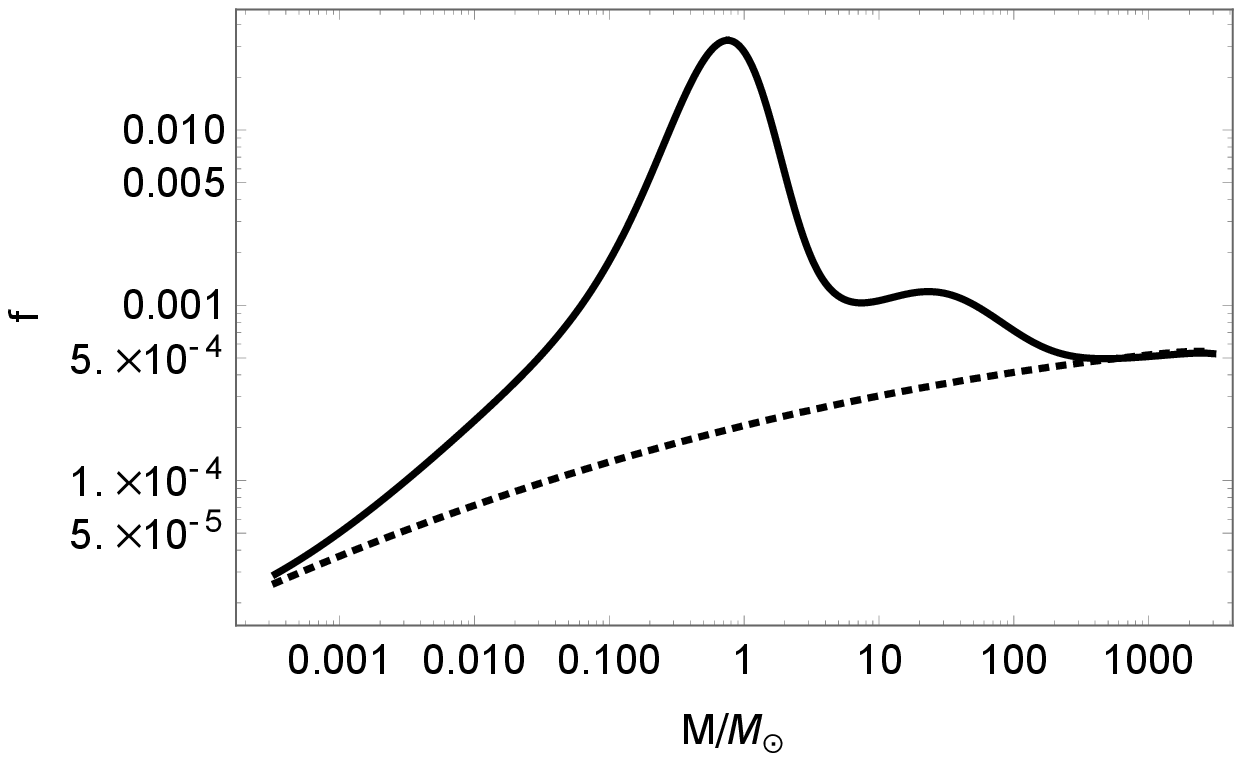}&
    \includegraphics[width=0.5\linewidth]{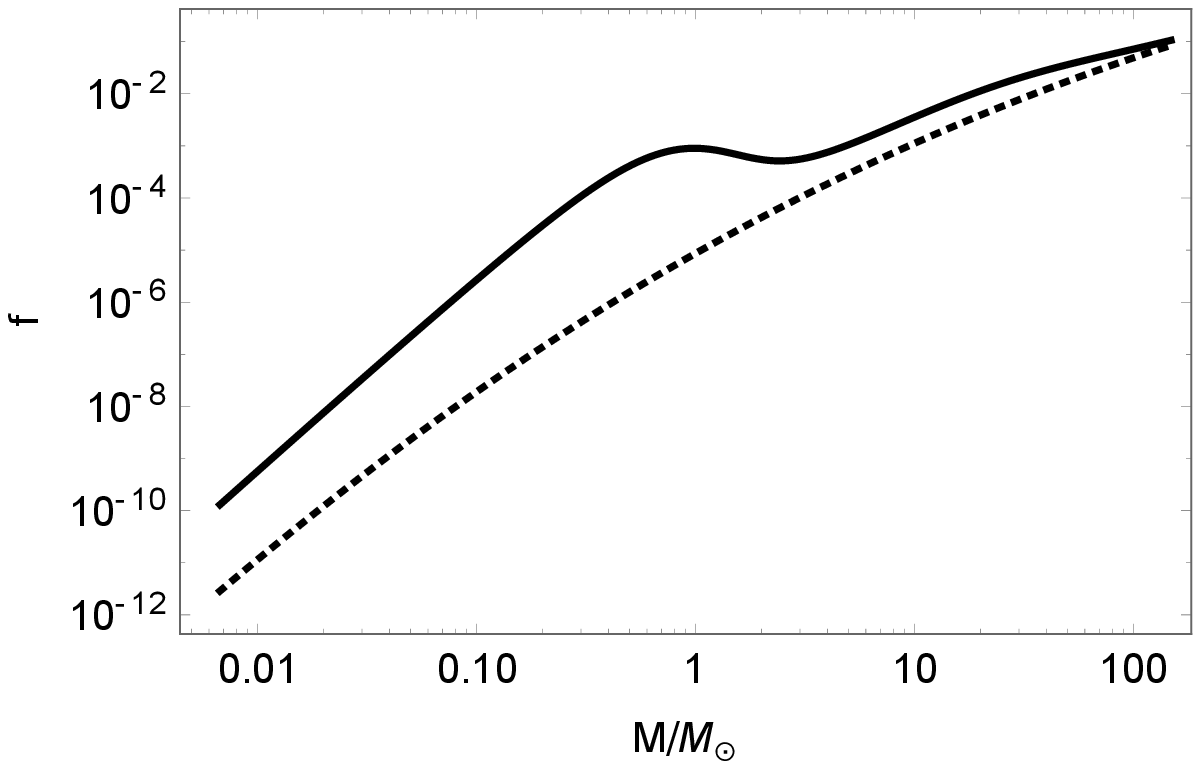}\\
\end{tabular}
\caption{\label{fig:f-tilted} The mass distribution of PBHs formed during the phase transition for two different spectral indices. 
The plot on the left shows $\sigma^2\propto \Mhub^{0.025}$ approximately corresponding to a spectral index of the primordial curvature perturbation of $n_s-1=-0.05$; while the plot on the right shows $\sigma^2\propto M^{0.1}$ ($n_s-1=-0.2$). Both power spectra have been normalised to $\sigma^2=0.004$ at $M=10 M_\odot$. The solid line is calculated using a time averaged value of $\delta_c$ while the dashed line assumes no phase transition and a constant critical value for collapse given by $\delta_c=0.453$.}
\end{figure*}

In order to provide some intuition about the effect of a scale dependent density contrast (measured at horizon entry) in Fig.~\ref{fig:f-tilted} we plot $f(M)$ for two different values of the spectral index $n_M$
\begin{equation}
\sigma^2(\Mhub)=0.004 \left(\frac{\Mhub}{10 M_\odot}\right)^{n_M},
\end{equation}
in both cases taking the same normalisation at $10 M_\odot$. The scale-dependence is related to the usual spectral index of the primordial curvature perturbation by $n_s-1\simeq -2 n_M$ because $\Mhub\propto a^2$ and $k=aH\propto 1/a$ during radiation domination.\footnote{The relation $n_s-1\simeq -2 n_M$ is not exact because the equation of state varies during the phase transition, and the conversion between the density contrast and the curvature perturbation depend on this. Therefore a scale invariant value of $\sigma^2$ does not exactly correspond to a scale independent primordial curvature perturbation.}

The left hand plot has a gentle scale dependence corresponding to $n_M=0.025$ ($n_s-1=-0.05$) which shows a clear peak in $f$ centred just below one solar mass. The peak has become relatively insignificant on the right hand plot which has a four times stronger scale dependence of $n_M=0.1$ ($n_s-1=-0.2$). 
In this case the heaviest PBHs will dominate unless the power spectrum is cut off on scales corresponding to a horizon mass between about 1 and 4 solar masses. 
In both of the plots shown the effect of the QCD phase transition is still very important, with the value of $f$ boosted by about a factor of 200 at $M=M_\odot$ compared to a calculation which neglects the reduction in the equation of state during this transition (compare the solid and dashed lines in Fig.~\ref{fig:f-tilted}). 

\subsection{Requirements for inflation to produce PBH with solar mass}

On large scales ($k\lesssim \mpc^{-1}$), the amplitude of the primordial curvature power spectrum is well measured by CMB and LSS observations to be around $10^{-9}$. On the much smaller scales associated with PBH formation, the power spectrum needs to grow by at least 6 orders of magnitude in order for any to form. The scale associated to the
PBHs detected by LIGO with $M\sim 30 M_\odot$ is $k \sim 10^6 \mpc^{-1}$ (see \eqref{M-k}) 
and such a rapid growth of the power spectrum requires a break down in slow-roll somewhere in between these scales \cite{Motohashi:2017kbs}. It does not however require that slow-roll is violated while the relevant scales producing PBHs are exiting the horizon, and hence the approximation of a constant spectral index (which is not related to the observed value of the spectral index on CMB scales) over this narrow range of scales may, in principle, still be appropriate. 
The breakdown in slow-roll may produce a peak of its own in the PBH mass distribution, although it is unclear how narrow this peak can be in a realistic model.
The growth in $f$ by 2 to 3 orders of magnitude for solar mass PBHs compared to the preferred LIGO mass range of $\sim30 M_\odot$ BHs poses a challenge for inflationary model building which seek to explain the LIGO BH mass range without contradicting any of the constraints on lighter PBHs. The softening in the equation of state during the QCD phase transition should be taken into account, and this will require a significantly more sharply peaked power spectrum than otherwise required when neglecting the effect of the QCD phase transition. To the best of our knowledge, none of the papers that construct an inflationary model capable of generating $\sim30 M_\odot$ black holes has taken this phase transition into account.  
Papers which place an inflection point or other feature at a small scale in an attempt to generate an interesting number of PBHs include 
\cite{GarciaBellido:1996qt,Clesse:2015wea,Garcia-Bellido:2017mdw,Kannike:2017bxn,Ezquiaga:2017fvi,Motohashi:2017kbs,Germani:2017bcs,Pattison:2017mbe,Ballesteros:2017fsr,Hertzberg:2017dkh,Pi:2017gih}. 

The power spectrum constraints on length scales smaller than $k\sim \mpc^{-1}$ 
are more model dependent, but CMB spectral distortion constrains the primordial power spectrum to be an order of magnitude too small on scales $k\lesssim 10^5 \mpc^{-1}$ \cite{Chluba:2012we} corresponding to a horizon mass of around $10^3 M_\odot$, meaning the power spectrum must grow significantly on scales between $10^5\mpc^{-1}$ and $10^6\;\mpc^{-1}$ in order for LIGO mass PBHs to be generated and the spectral distortion constraints to be evaded. Other small scale constraints include ultracompact minihalos,\footnote{It has been claimed that 
ultracompact mini halos already rule out the power spectrum growing so large on scales of $10^6 \; \mpc^{-1}$ \cite{Bringmann:2011ut}. However, these constraints are too strong because realistic halo formation is not as isolated as previously assumed leading to less steep density profiles \cite{Gosenca:2017ybi,Delos:2017thv} -- see also \cite{Nakama:2017qac}.} gravitational wave constraints from the pulsar timing array 
\cite{Saito:2008jc,*Saito:2009jt,Nakama:2016enz,Inomata:2016rbd,Orlofsky:2016vbd,Nakama:2016gzw} (for a recent review see \cite{Caprini:2018mtu}), big bang nucleosynthesis \cite{Barrow:2018yyg} and the dispersion of type 1a SNe  brightness due to lensing \cite{Ben-Dayan:2013eza,Zumalacarregui:2017qqd}.

For a slightly red spectral index we may tune the power spectrum such that only PBHs with masses comparable to the horizon mass during the QCD phase transition form in significant numbers, without requiring any additional cut-off or features of the power  spectrum on comparable scales. 
 Unless the primordial power spectrum has a sufficiently red tilt on small scales, the lightest PBHs will dominate, meaning that there then also needs to be a reduction in the power spectrum amplitude on even smaller scales than the ones we study in this paper.

Almost all of the constraints listed in the paragraph above assume a Gaussian distribution for the perturbations and may not apply if the perturbations are non-Gaussian, see e.g.~\cite{Nakama:2016gzw,Ando:2017veq}. Because PBHs form from extreme fluctuations deep in the tail of the probability density function, even relatively small amounts of non-Gaussianity can have an exponentially large effect on the formation rate of PBHs. With large non-Gaussianity, the amplitude of the primordial power spectrum can change by over an order of magnitude while keeping the number density of PBHs constant, see e.g.~\cite{Byrnes:2012yx,Young:2013oia}. However, if there is a non-negligible contribution of ``squeezed-limit'' non-Gaussianity generating a mode coupling between the long wavelength modes observable with the CMB and the smaller scales relevant to PBH formation then long-wavelength isocurvature perturbations will be generated  and the model is ruled out \cite{Tada:2015noa,Young:2015kda}.

\section{Summary}\label{sec:conclusions}

In this paper we have re-examined primordial black hole formation using an accurate Standard Model equation of state near the QCD phase transition, derived from lattice results \cite{Borsanyi:2016ksw}.  The QCD phase transition occurs at a time when the horizon mass is around 1 solar mass, and the softening of the equation of state as confinement sets in suggests that the Universe may be populated by PBHs with a mass distribution peaked at around a solar mass \cite{Jedamzik:1996mr}. 
There is some evidence for the existence of a non-stellar population of massive compact halo objects (MACHOs) of around a solar mass, and in addition, there has been much discussion that the merging black holes observed by LIGO (with masses of tens of solar masses) may have been primordial in origin. 

The density of PBHs depends sensitively on how the critical density perturbation $\deCrit$ is affected by the changing equation of state during the collapse process.  In the absence of a full collapse simulation, we explore the bounding assumptions that $\deCrit$ is either its value at horizon crossing or at the turn-around time for a spherically symmetric Gaussian initial perturbation. We also present time-averaged values, both linear and logarithmic, as intermediate cases.

The PBH mass distribution depends on the underlying density power spectrum.  
If the spectrum is scale-invariant in the range $10^{-3} \lesssim M_H/\Msun \lesssim 10^3\;$, 
we find that the peak mass 
is $0.7\;\Msun$ in the averaged cases, and bounded by $0.48  < \Mpeak/\Msun < 1.4 $. 
In the averaged case the FWHM is $0.30  < M/\Msun < 1.4 $. We emphasise that the main uncertainty is due to the crude modelling of the collapse dynamics, rather than in the QCD thermodynamics, motivating a more accurate study along the lines of Ref.~\cite{Musco:2012au}.
The boost to the PBH mass fraction caused by the softening of the equation of state cannot be neglected 
within the mass range $0.1 \lesssim M/\Msun < 100$; hence this is another reason why
a power-law PBH mass distribution in this range is quite unrealistic.  

Again in the scale-invariant case, the peak of the PBH mass density distribution is boosted by a factor at least one hundred over a universe with a pure radiation equation of state. A smaller boost extends over a wide range of masses $\mathcal{O}(0.1M_\odot)$ up to $\mathcal{O}(100M_\odot)$, with LIGO precursors ($M \sim 10 \Msun$) boosted by a factor of about 5.
If the fractional mass density in the range $10 < M/\Msun < 50$ is tuned consistent with the merger rate inferred from the LIGO observations,  $f \simeq 0.001$ \cite{Nakamura:1997sm,Sasaki:2016jop,Ali-Haimoud:2017rtz} (see also \cite{Raidal:2017mfl} which does not assume a monochromatic PBH mass spectrum, 
and \cite{Magee:2017vkk} who reached a different conclusion about the required mass fraction), 
a much larger density in solar mass black holes is found; PBHs make up 13\% of DM in the mass range $0.2 < M/\Msun < 1$. 
This fraction can be adjusted by small changes in the tilt of the density power spectrum. 

Therefore, a model with a large but quasi-scale-invariant primordial density power spectrum in the mass range 
{O}(0.1) $M_\odot$ up to {O}(100) $M_\odot$ can simultaneously account for both the LIGO black holes and potential MACHO observations in terms of PBHs. 
A detailed calculation of the merger rates that would arise from the PBH mass distributions calculated here is left for future study. 

Our results and formalism could easily be extended to calculate the PBH formation rate during any other phase transition during the early Universe. 
A generic expectation is that there will be a strong local peak in the formation rate whenever the equation of state parameter is reduced.
For example, a phase transition with critical temperature in the range $10$ TeV to $1000$ TeV could produce a peak in the PBH mass distribution in the 
range $10^{-10}\Msun$ to $10^{-14}\Msun$, where the constraints are weakest \cite{Carr:2009jm,Green:2014faa,Sasaki:2018dmp}.

\section{Acknowledgements}

CB was supported by a Royal Society University Research Fellowship. 
Mark Hindmarsh (ORCID ID 0000-0002-9307-437X) acknowledges
support from the Science and Technology
Facilities Council (grant numbers ST/L000504/1 and ST/P000819/1). We thank Guillermo Ballesteros for making a careful reading of the first version of this paper.

\appendix

\section{Microlensing constraints and solar mass PBHs}\label{sec:observations}

The debate on whether the dark matter is formed of elementary particles or compact bodies 
has been going on for some time (see e.g.~\cite{Trimble:1987ee} for an early review).
Any type of compact body making up the dark matter must be
non-baryonic, and the most plausible compact body candidate appears to
be primordial black holes.  
Microlensing observations provide one of the best studied and 
most competitive constraints on solar mass PBHs.


An early proposal for an observational signal of primordial black holes 
was that they should microlense distant quasars \cite{h93}. 
The situation then changed dramatically with results from an experiment to
detect compact bodies in the Galactic halo as they microlensed distant
stars in the Magellanic Clouds \cite{Alcock:2000ph}.  Some 15 events were observed,
consistent with a lens mass of around $0.5 M_{\odot}$, and at a higher
rate than could be accounted for by stars in the Galactic halo.
Alcock et al. \cite{Alcock:2000ph} also concluded that for their preferred halo model the lenses
would make up less than 50\% of the dark matter mass, their best estimate
being 20\%.  At the time this was seen as essentially ruling out compact
bodies of around a solar mass as dark matter candidates.  In a subsequent
re-analysis of their survey results, Alcock et al.~\cite{Alcock:2001a01} extended their limits
to include objects up to $30 M_{\odot}$ and concluded that this did not
alter their limits on the population of compact bodies in the Galactic
halo.

Other microlensing experiments in the Magellanic Clouds
\cite{Tisserand:2007t07,Wyrzykowski:2010mh}
produced conflicting results.  The objects detected by the MACHO
collaboration remained unexplained, which suggested that the deficit in
the number of events for a halo comprised of compact bodies might be
explained by incorrect assumptions about the structure of the Galactic
halo, and problems with estimating the detection efficiency for
microlensing events.  Recent observations implying a much lighter halo
than that assumed by the microlensing experiments \cite{Alcock:2000ph,Tisserand:2007t07,Wyrzykowski:2010mh}
prompted a re-examination of the assumptions upon which the analysis
of these groups was based \cite{Hawkins:2015uja}, with the conclusion that an
all MACHO Galactic halo with masses around a solar mass is not excluded.
The effect of changing the velocity dispersion of the halo model has
recently been explored by Green \cite{Green:2017qoa}, who found that
constraints on the mass fraction varied by almost an order of magnitude
between the halo models considered.   It has also recently been shown
\cite{Green:2017qoa,Calcino:2018cgd} that the effect of spatial clustering of
lenses or of broadening their mass function can effectively reduce or even
remove microlensing constraints on compact bodies in the Galactic halo.
To summarise, the constraints on compact bodies from microlensing rely on
a number of assumptions including the shape of the Galactic rotation curve,
the velocity dispersion in the halo, the clustering properties of the lenses,
and the efficiency with which the microlensing events are detected.  In view
of these uncertainties we feel that it is premature to rule out a population
of stellar mass black holes on the basis of microlensing studies in the
Galactic halo.

There have also been searches for `pixel-lensing' in M31 where unresolved
stars are being microlensed causing a pixel to brighten.  About 30
possible microlensing events have been detected \cite{Novati:2010c10}, but
there is still some debate as to the extent to which self-lensing distorts
the statistics \cite{Novati:2011ii}.   There does however seem to be a
good case for around 25\% of the M31 halo to be in the form of
compact bodies \cite{Novati:2005c05,Jong:2006d06}.  This assumes that the
the standard halo model used by the MACHO collaboration for the Milky Way
halo is also correct for M31, and so the issues raised in the previous
paragraph will equally apply to the pixel lensing results.

Another approach involved putting limits on any cosmological population of
compact bodies by observing the extent to which quasar images in
gravitationally lensed systems are being microlensed \cite{Mediavilla:2009m09,Pooley:2011gq}.
The observation of microlensing is inferred when the individual images of
a quasar system vary in brightness independently.  It was concluded that
their results only supported a small population of compact bodies, but
again the results depend on a number of uncertain assumptions.  These
include the adoption of a Single Isothermal Sphere model for the galaxy
halos along the line of sight to the the quasars despite the extensive
evidence favouring NFW profiles, and the uncertain contribution from the
distribution of the stellar population.  Sizes for the quasar accretion
discs and the extent to which the emission line regions can be microlensed
must also be assumed.

Predictions of the effect of microlensing on the brightness of Type Ia
supernovae have also been used to set bounds on any population of primordial
black holes \cite{Zumalacarregui:2017qqd}.  The authors make the case that
the population of PBHs is limited to around 30\% of the dark matter on the
basis of a lack of highly magnified lightcurves in samples of supernovae.
This result has been challenged by Garc\'{i}a-Bellido et al.~\cite{Garcia-Bellido:2017imq}  In particular, they query the assumptions for the size of the supernova
discs and choice of mass spectrum for the PBH masses.

Limits on the halo fraction of more massive ($\gtrsim 10 M_\odot$) bodies
come from several different sources which can be broadly divided into
dynamical and accretion constraints.  The effect of a population of MACHOs
on the distribution of wide binary star semiaxes in the Galactic halo has
been used to put limits on the halo fraction for perturbers of a range
of masses \cite{Monroy:2014m14}.  Their results are very sensitive to the
samples of binaries they choose, but broadly exclude a halo composed of compact
bodies with $M \gtrsim 20 M_\odot$.  Another approach to deriving a
dynamical constraint on the population of MACHOs comes from the
distribution of stars in dwarf galaxies.  It is argued that dark matter in
the form of massive compact bodies will result in mass segregation in the
stellar population leading to a depletion of stars in the centre of the
galaxy \cite{Brandt:2016b16,Koushiappas:2017k17}. This is not seen in the
observed stellar profile, implying a limit of $M \gtrsim 10 M_\odot$ for an
all MACHO halo.  There are however a number of uncertainties in this procedure.
These are discussed by the authors and include the velocity distribution
of the dark matter particles, the uniformity of the dark matter density, the
relative velocity dispersion of stars and dark matter and the dark matter
profile.  Perhaps most importantly, they assume that there is no Intermediate
Mass Black Hole at the centre of the galaxy.  As Lee et al. \cite{Lee:l17}
and Clesse \& Garcia-Bellido \cite{Clesse:2017bsw} point out, the presence of
an IMBH could negate their constraints.

The accretion constraints on the PBH population come firstly from the
predicted interaction of PBHs with the inter-stellar medium, which would
result in an excess of X-ray sources.  This leads to a predicted X-ray
luminosity function for a given population of PBHs which can then be
compared to the observed luminosity function \cite{Inoue:2017i17}, or
cross-correlated with a radio source catalogue \cite{Gaggero:2017g17}.  This approach appears to
excludes PBH dominated dark matter in the Milky Way for
$M \gtrsim 10 M_{\odot}$. 
An alternative approach is to look for distortions in the
Cosmic Microwave Background resulting from accretion onto PBHs in the
early universe.  There has been considerable divergence in the results
from this procedure, reflecting the complications of the accretion physics
involved.  The results put constraints on PBHs with mass
$M \gtrsim 5 M_{\odot}$ \cite{Aloni:2017a17a} to $M \gtrsim 100 M_{\odot}$
\cite{Ali:2017a17b} as the dominant component of dark matter.  However, as
both these groups point out, some aspects of the analysis are still
uncertain.

The observational case for compact bodies as dark matter has been
summarised by Hawkins \cite{Hawkins:2011h11}, where it was argued that there is
good reason to believe that the light from most, if not all, quasars is being
microlensed by compact bodies of around a solar mass, in which case the
population of lenses will have an optical depth to microlensing close
to unity.  This implies \cite{Press:1973p73} that the lenses could make up an
appreciable fraction of the cosmological dark matter.  In view of the
uncertainties outlined in \cite{Hawkins:2015uja}, it is possible that they could even
constitute all of it.  Therefore, one cannot securely conclude from
microlensing observations that the PBH mass fraction $f(M)$ is much less
than 1 at around a solar mass.

\bibliographystyle{JHEP} 

\bibliography{PBH-bibliography}

\end{document}